# On "Inconsistencies of metalens performance and comparison with conventional diffractive optics"

Amir Arbabi[1,*] & Andrei Faraon[2]

[1]Department of Electrical and Computer Engineering, University of Massachusetts Amherst, Amherst, MA, USA
[2]T. J. Watson Laboratory of Applied Physics and Kavli Nanoscience Institute, California Institute of Technology, Pasadena, CA, USA
[*]e-mail: arbabi@umass.edu

**It was recently claimed[1] that reported focusing efficiency values of high numerical aperture metalenses are inconsistent with a theoretical bound, and their measurement results are incorrectly interpreted. We review the article and conclude that these claims are not well supported.**

Optical metalenses, flat lenses made of scatterers, have gained attention for their ultrathin form, potential low cost, high numerical aperture (NA), focusing with high efficiency, and multifunctionality[2]. Menon and Sensale-Rodriguez recently questioned these features[1], claiming: 1. Reported focusing efficiency values exceed a theoretical bound, and measurement procedures are flawed, providing no reliable evidence of high focusing efficiency of high NA metalenses. 2. Metalenses offer no advantage over multi-level diffractive elements in multi-functionality. We show that their claims and conclusions are based on incorrect assumptions and mistakes.

Like other diffractive lenses, metalenses with multiple phase-wrapped zones focus light via diffraction, acting as gratings with varying periods and diffracting incident light into multiple orders (see Fig. 1a). Flat lens design aims to increase the power directed to the main focal point (+1 diffraction order, shown by solid green arrows in Fig. 1a). The fraction of incident power that is directed toward the main focal point is referred to as the (absolute) focusing efficiency[2] and is used to compare flat lens designs. To measure the focusing efficiency, a circular aperture is placed in the lens focal plane, blocking unfocused light while allowing the main focal spot's power to pass through.

Many metalenses are designed using a unit-cell-based approach, where the scatterers are arranged on a periodic lattice, and their dimensions and shapes are chosen based on the desired local optical transformation. Using this approach, we demonstrated high focusing efficiency of high NA amorphous silicon metalenses via simulations and experiments[3]. Subsequently, others demonstrated similar high-NA metalenses and reported high focusing efficiencies.

More recently, Chung and Miller introduced a theoretical bound on the focusing efficiency of unit-cell-based metalenses[4]. Menon and Sensale-Rodriguez noted several high-NA metalenses' measured efficiencies exceed this bound, attributing inconsistencies to using large-radius apertures in their measurements[1]. We examined the theoretical bound of Ref.[4] and found a mistake in its derivation. The corrected bound aligns with the reported experimental results, resolving the claimed inconsistency.

Menon and Sensale-Rodriguez claimed that focusing efficiency values are overestimated when the aperture radius is large compared to the full-width at half-maximum (FWHM) of the focal spot. By determining the fraction of an Airy pattern power inside a circular aperture, they argued that large apertures lead to overestimation. However, their analysis contained mistakes: using an incorrect Airy pattern expression (assuming $\mathrm{NA} = \tan\theta$ instead of $\sin\theta$), erroneous computation of encircled power, assuming the high-NA flat lens focal spot is an Airy pattern, and misusing aperture radius vs. FWHM spot size as the focusing efficiency error metric.

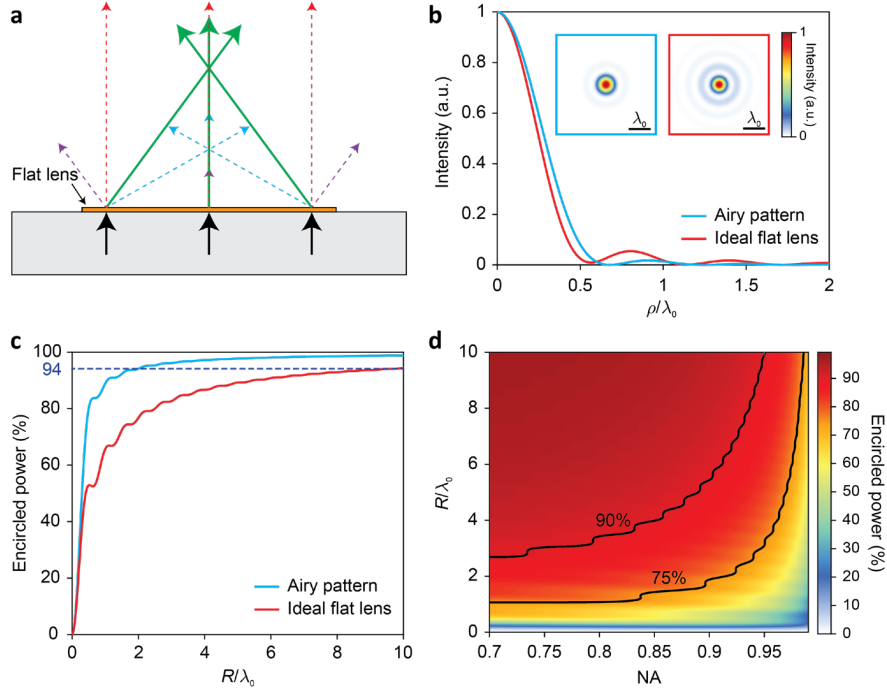

**Fig. 1. Focusing characteristics of flat lenses. a,** A flat lens has multiple focal points corresponding to different local diffraction orders. 0$^{th}$ order (red dashed arrows), +1 order (main focal point, solid green arrows), -1 order (dashed purple arrows), and +2 (dashed blue arrows) are shown as examples. **b,** Radial cuts of the Airy pattern and the focal spot intensity of an ideal flat lens with an NA of 0.9 that is uniformly illuminated with unpolarized light. The insets show the corresponding 2D plots. **c,** The percentages of the power of the Airy pattern and the focal spot of an ideal flat lens with NA of 0.9 that are confined within a circle with a radius of $R$. **d,** The percentage of the power of the focal spot of an ideal flat lens that is confined within a circle with a radius of $R$ for different NA values. 75% and 90% contours are shown. $\lambda_0$: vacuum wavelength.

The focal spots of uniformly illuminated aplanatic lenses are well approximated by the Airy pattern[6]. However, single-layer flat lenses are not aplanatic[7], and the focal spot of a high-NA ideal flat lens deviates significantly from an Airy pattern[8]. We refer to a local flat lens with 100% focusing efficiency that is corrected for spherical aberration as an ideal flat lens. Figure 1b shows the Airy pattern and the focal spot of an ideal flat lens with an NA of 0.9 when uniformly illuminated by normally incident unpolarized light[8]. As Fig. 1b shows, the intensities of the outer rings are larger in the ideal flat lens' spot than the Airy pattern. Therefore, a larger radius is needed to obtain the same encircled power percentage for a flat lens than for an aplanatic lens (Fig. 1c). For example, for the NA of 0.9, 94% of the Airy pattern's power passes through an aperture with a radius of $2\lambda_0$ ($\lambda_0$: vacuum wavelength), but a radius of $9.5\lambda_0$ is needed for the same percentage in an ideal flat lens.

Figure 1d shows the encircled power for uniformly illuminated ideal flat lenses, indicating that the aperture radius should increase with NA to maintain the same encircled power percentage. However, a large aperture radius can overestimate the focusing efficiency by allowing unfocused light to pass through. An upper bound on the fraction of unfocused power passing through the aperture is the ratio of the aperture area to the flat lens area since the unfocused power is distributed over an area larger than the lens (Fig. 1a). To determine overestimation, the aperture area should be compared to the lens area, not its focal spot.

Menon and Sensale-Rodriguez claim gross overestimation of focusing efficiency in Ref.[5], where an aperture with a radius 8.5 times the FWHM focal spot size was used. The metalens in Ref.[5] has a radius of 200 $\mu$m, an NA of 0.78, operates at $\lambda_0 = 0.915$ $\mu$m, and a 5-$\mu$m-radius aperture was used in its measurement. Using Fig. 1d data and the upper bound on unfocused power, we find 94% of focused power and less than 0.1% of unfocused power pass through the 5-$\mu$m-radius aperture. Thus, the reported 77% focusing efficiency is an underestimation, invalidating the claim that high focusing efficiency values of high-NA metalenses are not backed by rigorous simulation or experimental results. In fact, a full-wave simulation is also presented in Ref.[5] and the results are consistent with measured values.

Their other claim that metalenses offer no clear advantage over conventional diffractive elements in multi-functionality is also unsupported. Compared to multi-level diffractive optics, metalenses provide higher efficiency, higher spatial resolution, and simpler fabrication. For example, the device in Ref.[9], which is cited as evidence of multifunctional multi-level elements, has the highest efficiency of 25%, ~$20\lambda_0$ spatial resolution, and require complex fabrication (four-level deep ion milling etches into two birefringent crystals followed by their alignment and bonding[9]). In contrast, polarization multifunctional metalenses are over 90% efficient, offer a spatial resolution of $\lambda_0/2$, and are fabricated using a single-step[10]. This demonstrates that metalenses indeed offer significant advantages in terms of multifunctionality.